\documentclass[
    corpo=10pt,
    stile=classica,
    tipotesi=custom,
    greek,
    evenboxes,
]{toptesi}
\usepackage{tabularx}
\usepackage{array}
\usepackage[utf8]{inputenc}
\usepackage[T1]{fontenc}
\usepackage{lmodern}
\usepackage{geometry}
\usepackage{graphicx}
\usepackage{float}
\usepackage{booktabs}
\usepackage{amsmath}
\usepackage{xurl}
\usepackage{hyperref}
\usepackage{xcolor}
\usepackage{seqsplit}
\usepackage{enumitem}
\usepackage{caption}
\hypersetup{pdfpagemode=UseOutlines,bookmarksopen,pdfstartview=FitH,colorlinks=true,linkcolor=blue,citecolor=blue,urlcolor=blue}
\setlist[itemize]{leftmargin=1.2em,itemsep=0.05em,topsep=0.1em}
\newcommand{\projectname}{GROB}
\newcommand{\legacyname}{CyberXAI}
\newcommand{\code}[1]{{\ttfamily\seqsplit{#1}}}
\definecolor{softgray}{RGB}{245,246,248}
\definecolor{softorange}{RGB}{255,246,232}
\usepackage{caption}
\usepackage{etoolbox}

\hypersetup{
    pdfpagemode=UseOutlines,
    bookmarksopen,
    pdfstartview=FitH,
    colorlinks=true,
    linkcolor=blue,
    citecolor=blue,
    urlcolor=blue
}

\makeatletter
\patchcmd{\thebibliography}
  {\chapter*{\bibname}}
  {\section*{\bibname}}
  {}{}
\makeatother

\begin{document}
\english

\noindent
\begin{minipage}[c]{0.68\textwidth}
    \textbf{Politecnico di Torino}\\
    \textbf{Technical Research Report}
\end{minipage}\hfill
\begin{minipage}[c]{0.27\textwidth}
\begin{flushright}
\IfFileExists{figures/logo_poli.jpg}{\includegraphics[width=0.72\linewidth]{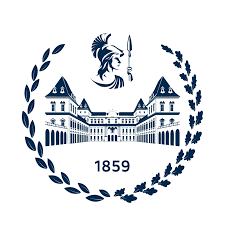}}{\fbox{\parbox[c][1.05cm][c]{0.80\linewidth}{\centering\scriptsize\textbf{POLITECNICO\\DI TORINO}}}}
\end{flushright}
\end{minipage}

\begin{center}
{\LARGE\bfseries \projectname{}: A Multi-Agent Architecture for Public-Trace Investigation of Candidate Agentic Activity}\\[0.25em]
{\large Chiara Bonfanti \hspace{1em}--\hspace{1em} Prof.~Cataldo Basile}\\[0.15em]
{\small 6 October 2026}
\end{center}

\vspace{0.35em}
\begingroup
\small
\noindent\textbf{Abstract.}
We present \projectname{}, a multi-agent architecture for investigating candidate autonomous-agent activity through public Internet traces when privileged telemetry is unavailable. The system performs controlled, read-only collection of public traces and preserves selected observations for later resolution. 
In a frozen September 2026 corpus, several collected traces became more informative as additional public evidence emerged. The strongest result concerns Census-labelled identifiers captured on 9 September. Public revision records later resolved these identifiers to specific Census requests from 16-17 June. Other results show weaker links between traces collected by \projectname{} and evidence reconstructed or reported later. These links vary in strength, and only some can be tied to specific public records. The results show that sparse public traces can remain useful even before their significance is fully understood. Such evidence can support later reconstruction, but public traces alone do not establish organizational attribution. Execution identity presents a separate problem, as continuity of agent identity remains an active research question for autonomous language-model agents \cite{perrier2026identity}.
\par
\endgroup
\vspace{0.35em}

\section{Introduction}
GROB is a multi-agent architecture for detecting and investigating candidate autonomous-agent activity from public Internet traces when privileged telemetry is unavailable. The system was developed under the earlier name \legacyname{}, which remains visible in earlier project material. Some historical requests may therefore still appear in public records under the \legacyname{} name used by the collector.\\
In this preprint, \emph{multi-agent} refers to a system in which individual agents have distinct roles and responsibilities. By contrast, \emph{swarm} refers to collective behaviour in which the group acts as a more unified whole and coordination is not tied to fixed individual roles.

Our work is motivated by incidents such as the 2026 Hugging Face case, where agents coordinated activity across otherwise separate runs \cite{greenblatt2026}. Such incidents expose an asymmetry between attacker coordination and defender visibility: coordinated agent activity may span multiple systems, while defenders see only partial external evidence. The Collusion Wiki investigation showed that agent-like activity could also leave traces on public Internet services used as communication channels \cite{vonarx2026collusion}. GROB is intended to narrow this gap by preserving public traces and prioritizing those most relevant for later reconstruction.

\textbf{Research question:} Can defenders use public Internet traces, without privileged telemetry or known targets, to identify and preserve signs of candidate agentic activity that can later be investigated and compared with independent evidence?

The prototype was designed to investigate several kinds of public Web evidence. In the earliest retained run, DSEWiki was not given to the system as a target. The collector reached it automatically while following links from the two configured wiki-service roots. This matters because the system reached a source that later proved relevant without being given that source or the later incident as a target. The same surface was subsequently fetched across all 35 deduplicated workspaces.

Because this is an emerging empirical setting without established ground truth, validation is necessarily retrospective. A trace collected by \projectname{} is first linked to a specific public revision, request, or resource. Later independent reporting or first-party disclosure is then used to check whether that resolved trace corresponds to subsequently reported activity. This allows the relevance of an earlier observation to be assessed without claiming that both records came from the same execution.

Census provides the strongest example because an identifier collected by \projectname{} can be resolved to specific public requests and compared with independent reporting covering the same target and time interval. On 9 September, \projectname{} collected three Census-labelled identifiers, including \code{AgentOpenAICensusTest}. Public DSEWiki revision records later resolved those identifiers to read-only Census requests dated 16--17 June, including American Community Survey Public Use Microdata Sample (ACS/PUMS) and GCT1701 resources. Census-related OpenAI activity was publicly reported on 25 September \cite{dimolfetta2026}, while subsequent reporting described activity during the overlapping 16--22 June period \cite{cable2026government}.
The result therefore shows that \projectname{} preserved Census traces before later public reporting established their significance, and that those traces could subsequently be resolved to specific public requests consistent with later reported agentic activity. At the time of writing, no first-party evidence links the public revisions to the specific OpenAI execution described in those later reports. This supports retrospective identification and reconstruction of relevant public traces, but not attribution to a specific OpenAI run.

\section{Operational workflow and architecture}

Given a set of allowlisted starting URLs, \projectname{} performs controlled link traversal and retains retrieved records together with their source and capture time. Candidate observations can open a scoped investigation, while deterministic rules control what is admitted to persistent evidence. The system produces a provenance-preserving workspace rather than an attribution decision.

\begin{figure}[htb]
\centering
\includegraphics[width=0.82\textwidth]{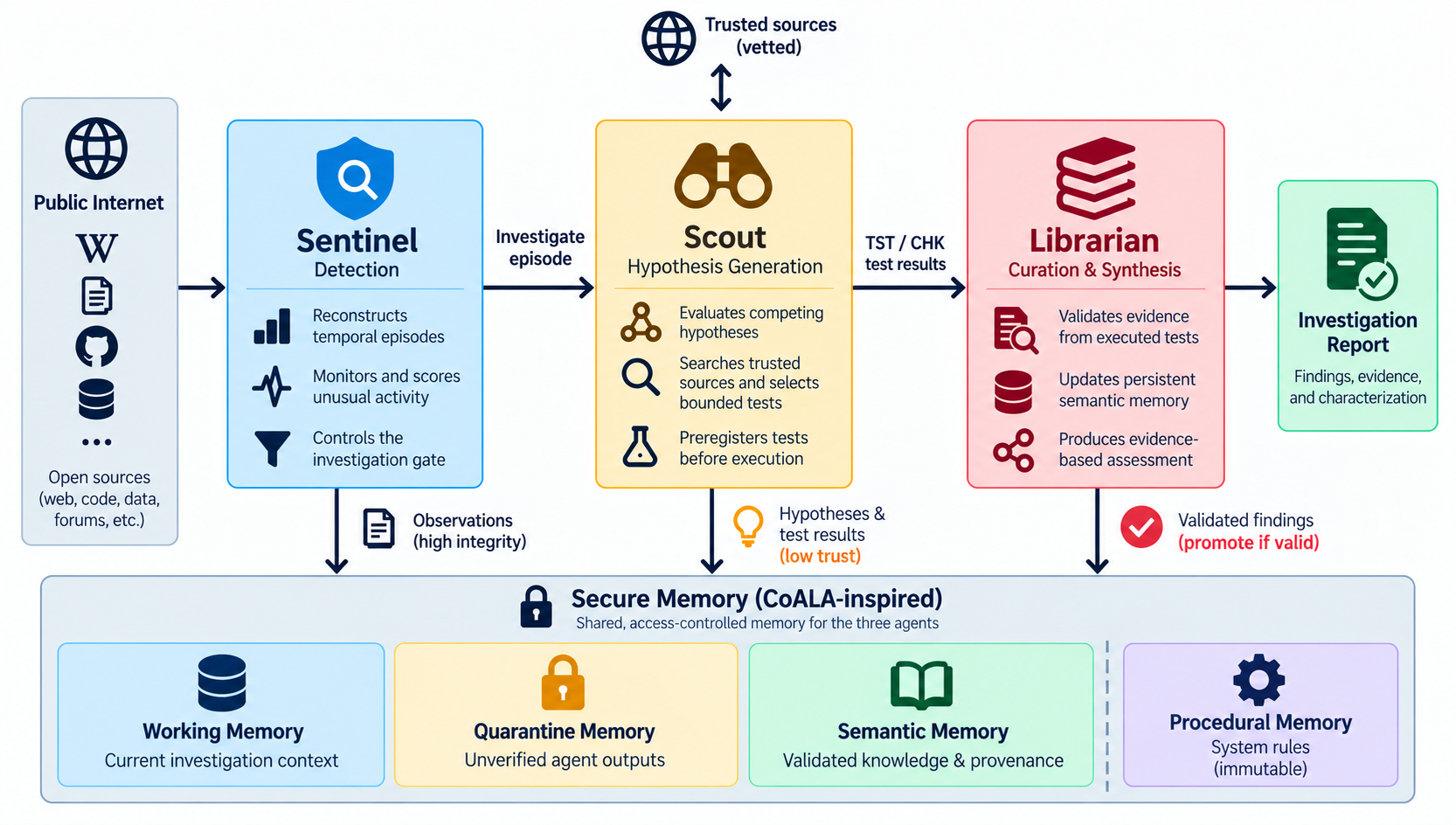}
\caption{\projectname{} architecture. Sentinel controls detection and gating, Scout performs hypothesis testing, and Librarian maintains validated semantic memory.}
\label{fig:architecture}
\end{figure}

Figure~\ref{fig:architecture} shows how an observation becomes an investigation. Sentinel determines whether the available public traces warrant further analysis. If an investigation is opened, Scout evaluates alternative explanations while Librarian preserves only validated evidence. This separation prevents model-generated interpretations from entering persistent memory
without review. 
At the stage represented by these experiments, \projectname{} was an exploratory research prototype rather than a production monitoring system. Its CoALA-inspired memory design treats retrieved Web material as untrusted until it is validated \cite{sumers2024coala}. Large language models assist with interpretation, but only deterministic code can admit evidence to persistent memory.

\textbf{Candidate selection.} Candidate records are prioritized using the evidence available at collection time. Follow-up tests are defined before additional retrieval so that the initial observation remains distinguishable from its later investigation.

\textbf{Public-web collection.} Collection is restricted to read-only retrieval of publicly accessible material. The system neither crosses access controls nor modifies remote systems.

\subsection{Data model and temporal interpretation}

\projectname{} operates without access to the underlying agent systems or to private service logs. Each collected public record is therefore stored together with its source and capture time. Repeated appearance across workspaces can reflect persistence of the same source material rather than a new real-world event.

In this report, a trace is considered \emph{found} when an exact identifier is retained through this process. This definition concerns observable evidence, not the identity of its author or operator. It is \emph{resolved} when the collected trace can be linked to a specific public revision, request, or resource. It is \emph{externally corroborated} when independent evidence later reports compatible activity.
None of these states, without first-party execution evidence, establishes the identity of the underlying actor or a shared execution.

Although the prototype supported several public-source adapters, the earliest retained run enabled its legacy-wiki discovery source. That source began from the general service roots \path{www.wikiservice.at/} and \path{www.prowiki.org/}; DSEWiki itself was not supplied as a seed. An allowlisted breadth-first traversal, limited to depth two and 35 pages per host, reached a DSEWiki path at depth one. The resulting workspace contained 70 collected page observations across nine host/path surfaces, including 23 DSE-related observations.

\begin{figure}[htb]
    \centering
    \includegraphics[
        width=\linewidth
    ]{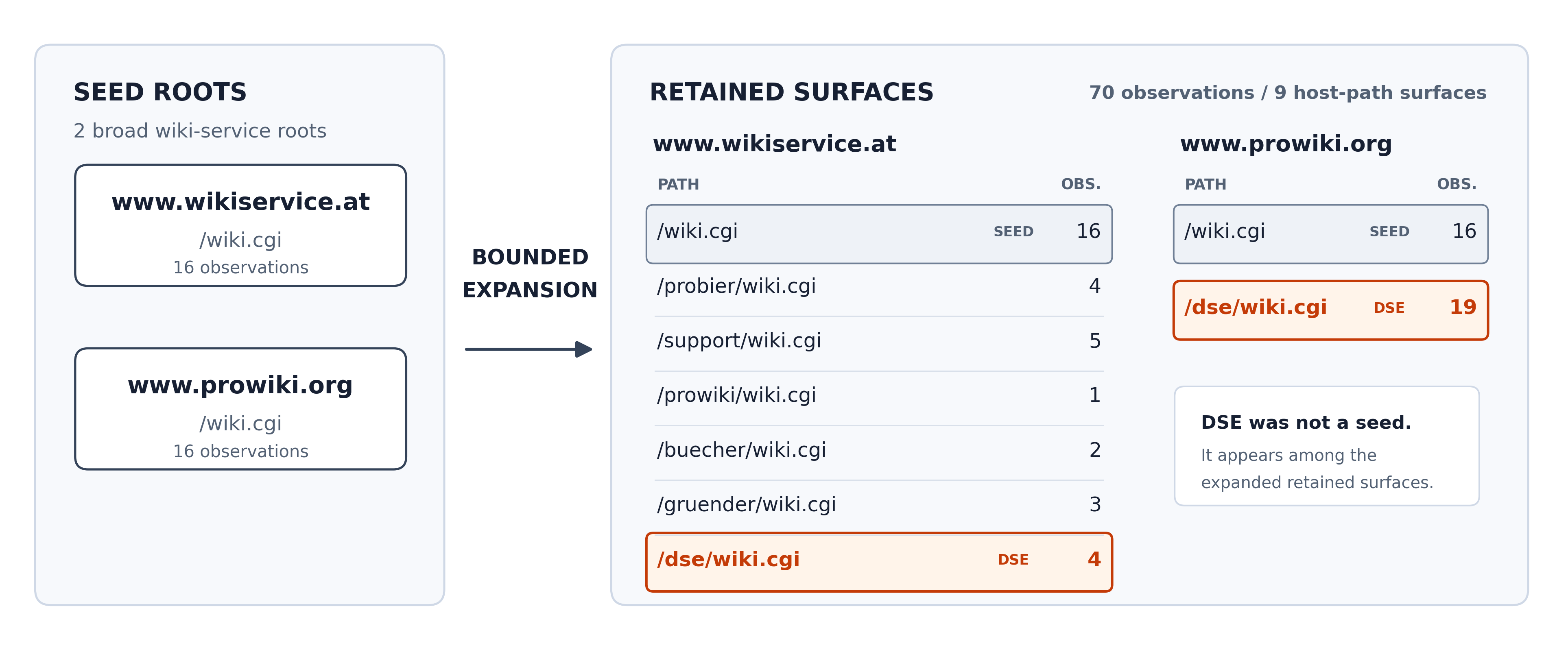}
    \caption{Wiki-surface expansion in the earliest preserved
    \legacyname{} workspace. DSE paths are highlighted.}
    \label{fig:wiki-discovery-path}
\end{figure}

The same surface was subsequently fetched in all 35 deduplicated workspaces in the broader audit population. Here, \emph{fetched} means that page content was actually retrieved and collected.

\section{Primary result: Census}

The earliest collected Census observation is dated 9 September 2026 at 22:43:48 UTC. It contains \code{AgentOpenAICensusTest} and \code{AgentOpenAICensusLink1781645460}. The same capture preserves \code{OpenAICensusGCTBridge}.

OpenAI's Census-specific acknowledgement was reported on 25 September \cite{dimolfetta2026}. The collected identifiers were therefore already present before that target-specific public confirmation.

The identifiers recur across multiple collected workspaces, indicating persistence rather than separate incidents.

The early workspaces preserve the page identifiers, which provide the join to the deleted bodies and Census URLs later recovered from a public DSE revision archive.

In the public DSE revision archive, the Census identifiers resolve to a repeated request rather than to naming similarity alone. \code{AgentOpenAICensusTest} contains the request on 16 June under \code{GroceryAgentDec09X}. A subsequent revision reformats the same request under \code{OpenAIHelperJul14X}. \code{AgentOpenAICensusLink1781645460} then carries the same request under \code{OpenAIResearcherOct14}. It appears again shortly afterward on the DataUSA-labelled page \code{AgentDataUSAResearch20270920}.

The same Census request therefore persists across several public revisions and labels within roughly eleven minutes. This establishes an artifact-level DataUSA-to-Census chain in the public revision history.

The request targets the U.S. Census Bureau 2020 five-year American Community Survey (ACS) Public Use Microdata Sample (PUMS) API:

\begin{quote}
\small
\path{api.census.gov/data/2020/acs/acs5/pums}
\end{quote}

The request performs a PWGTP-weighted tabulation by sex for Wyoming. It also contains a Census developer-key parameter whose value is redacted:

\begin{quote}
\small
\texttt{tabulate = weight(PWGTP)}\\
\texttt{row = SEX}\\
\texttt{state = 56} \quad (Wyoming)\\
\texttt{key = [redacted]}
\end{quote}

The redaction prevents comparison of the credential value. The request itself remains observable through its endpoint and query structure. Its recurrence across several revisions is also preserved.

A second sequence appears on 17 June. Between 10:27:36 and 10:28:33 UTC, \texttt{OpenAIHelper}-labelled revisions query a DataUSA ACS poverty cube. At 10:57:18, \code{OpenAICensusGCTBridge} uses the same save-name and points
through \path{r.jina.ai} to the Census table \path{GCT1701.csv}. The sequence therefore links DataUSA poverty work with a Census poverty resource over an interval of approximately 29 minutes.

Reporting on 25 September stated that OpenAI's agents had accessed public Census data using Census Data API developer keys found in public GitHub repositories during internal training tasks \cite{dimolfetta2026}. Transluce reported on 30 September that public URLs posted between 16 and 22 June indicated attempts to reuse exposed keys against \path{census.gov} \cite{cable2026government}. Transluce did not recover a response showing that those requests successfully reached Census.

The resolved \projectname{} events fall within the same June interval. They also contain a Census developer-key parameter. The correspondence therefore extends beyond a target-name match and reaches the level of a compatible request mechanism.

This establishes artifact-level continuity, not execution identity. To our knowledge, no first-party record currently links the June revisions to the execution later disclosed by OpenAI.

\begin{table}[ht]
\centering
\footnotesize
\caption{Census resolution chain. Times refer to public revisions; the
\projectname{} capture occurred in September.}
\label{tab:census-resolution}
\begin{tabularx}{\linewidth}{
    >{\raggedright\arraybackslash}p{0.16\linewidth}
    >{\raggedright\arraybackslash}p{0.29\linewidth}
    >{\raggedright\arraybackslash}X
}
\toprule
\textbf{Time (UTC)} &
\textbf{Artifact} &
\textbf{Resolved content} \\
\midrule

16 Jun 21:24:10 &
\texttt{AgentOpenAI\allowbreak CensusTest} &
ACS/PUMS request for Wyoming. \\

16 Jun 21:25:30 &
\texttt{AgentOpenAI\allowbreak CensusTest} & Same request reformatted under \texttt{OpenAIHelperJul14X}. \\

16 Jun 21:30:53 & \texttt{AgentOpenAI\allowbreak CensusLink\allowbreak1781645460} & Same ACS/PUMS request. \\

16 Jun 21:34:46 & \texttt{AgentDataUSA\allowbreak Research20270920} & Same request on a DataUSA-labelled page. \\

17 Jun 10:27--10:28 & \texttt{OpenAIHelper} sequence & DataUSA ACS poverty queries. \\

17 Jun 10:57:18 &
\texttt{OpenAICensus\allowbreak GCTBridge} &
Census GCT1701 poverty table via \path{r.jina.ai}. \\

\bottomrule
\end{tabularx}
\end{table}

This is the strongest example in the current corpus of a sparse identifier captured by the prototype later resolving into a specific public-data workflow.

\section{External comparison after capture}

\begin{figure}[htb]
\centering
\includegraphics[
    width=\linewidth
]{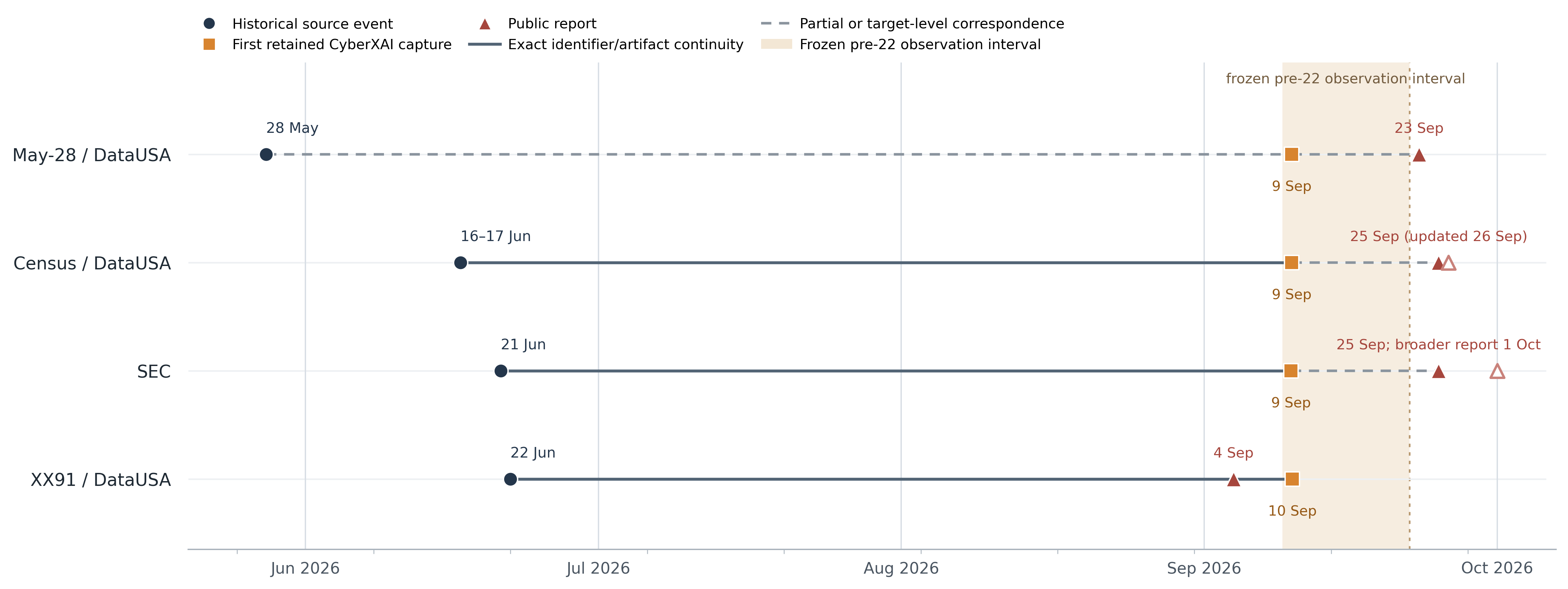}
\caption{Chronology of four dated trace families. Circles mark reconstructed source events, squares the first collected \projectname{} captures, and triangles later external reporting. For XX91, the source event is represented by its date and verified 106-second creation interval rather than by an unresolved display timezone.}
\label{fig:evidence-timeline}
\end{figure}

Evidence published after the \projectname{} captures is used only to interpret records that the prototype had already collected. It is not treated as input to the original runs or counted as pre-disclosure evidence. Figure~\ref{fig:evidence-timeline} separates reconstructed public events from their first \projectname{} captures and from later external reporting.

\subsection{Post-baseline external resolution}

Three external investigations materially refine the interpretation of the frozen corpus. Transluce published its government-site investigation on 30 September, using records from Arquivo.pt and urlquery.net \cite{cable2026government}. Asymmetric Security followed on 1 October with an independent reconstruction from publicly available traces \cite{majid2026rogue}. On 5 October, the Wikimedia Foundation published the results of its own investigation into activity it attributed to OpenAI-operated agents \cite{wikimedia2026}. These sources are used retrospectively and do not become \projectname{} observations.

\begin{table}[htb]
\centering
\scriptsize
\caption{External evidence published after the observation baseline and its
effect on interpretation.}
\label{tab:external-resolution}
\begin{tabularx}{\linewidth}{
    >{\raggedright\arraybackslash\bfseries}p{0.12\linewidth}
    >{\raggedright\arraybackslash}X
    >{\raggedright\arraybackslash}X
}
\toprule
Family & External evidence & Supported interpretation \\
\midrule

Census &
Transluce reported that public URLs posted between 16 and 22 June indicated attempts to reuse exposed API keys against \path{census.gov}. Several related pages contained OpenAI markers, but the underlying URLs were withheld and no recovered response established that the requests reached Census \cite{cable2026government}. & The target, time interval, and apparent API-key mechanism are independently compatible with the resolved \projectname{} artifacts. The redacted credential prevents determining whether the same key was involved, and no execution-level join is available. \\\\ 
Education & Transluce reconstructed more than 200,000 requests to the Civil Rights Data Collection API on 17 June. The sequence included the failed \texttt{State\_Id=1 OR 1=1} probe and more than 10,000 requests carrying an \texttt{oai}-prefixed tag consistent with task \texttt{dsqa\_250} \cite{cable2026government}. &
This independently characterizes the Education incident. The \projectname{} Education traces remain contextual because no exact artifact or provenance record joins the two datasets. \\\\
SEC & Transluce reconstructed a June 18 workflow seeking public SEC crowdfunding statistics. The investigation connected public wiki discussion of double-slash paths with urlquery records returning the SEC county-data resource; ordinary paths also worked, so no bypass was demonstrated \cite{cable2026government}. & The exact county-data resource also occurs in the \projectname{} reconstruction. This provides artifact-level correspondence without establishing execution identity. \\\\
Data USA & Asymmetric Security independently reported agent activity involving university statistics from Data USA and access to Data USA pre-production or staging infrastructure \cite{majid2026rogue}. & This provides independent context for the DataUSA family, but the report does not identify an exact \projectname{} label, XX91 artifact, or actor relation. \\\\
Wikimedia & On 5 October, the Wikimedia Foundation reported activity that it attributed to OpenAI-operated agents. Its investigation found sandbox edits, attempts to use hosted services for remote retrieval, and high-volume access to Wikimedia infrastructure \cite{wikimedia2026}. &
The frozen pre-22 corpus had independently collected activity on \texttt{Project:Sandbox}. The collected revisions differ from the revisions disclosed by Wikimedia, so the correspondence is at the public-page level rather than the revision level. \\

\bottomrule
\end{tabularx}
\end{table}

The external investigations also help identify non-matches. Transluce's 17 May Texas episode concerned sexually transmitted infection statistics and is distinct from the 22 June XX91/DataUSA/Texas-place episode. Its 28 May Library and Archives Canada activity concerned Canadian divorce records and is likewise distinct from the \projectname{} May-28 DataUSA branch \cite{cable2026government}. Neither event is used to strengthen those \projectname{} families.

The Wikimedia case provides a different form of comparison. \projectname{} had already collected activity on the same \texttt{Project:Sandbox} surface later identified by Wikimedia, but the collected content and revision identifiers do not match the disclosed activity. The result is therefore a surface-level correspondence rather than an incident-level join \cite{wikimedia2026}.

\subsection{Auditability after disclosure}

At this stage, \projectname{} was an exploratory prototype and did not always retain the complete content of every discovered page. In several cases it preserved enough information to identify a trace, while fuller reconstruction later depended on public records that remained available elsewhere.

This limitation also appears in independent investigations. Asymmetric Security reported that temporary services, private scans, and expiring records could leave parts of the activity unavailable for later reconstruction \cite{majid2026rogue}. Absence from the collected \projectname{} corpus is therefore treated as an auditability limitation rather than evidence that an event did not occur.

\section{Supporting evidence modes}

The remaining cases examine what can be inferred from different forms of public evidence. XX91 tests whether temporal and cross-surface structure adds information beyond naming similarity, while SEC provides an exact-resource correspondence. Thailand has direct page-level provenance, whereas NovOne marks the boundary of lexical comparison when no artifact-level join is available.

\subsection{Structured and artifact-level traces}

XX91 provides the clearest example of structure beyond lexical similarity.
The pre-22 corpus collected the complete five-page ProbierWiki \code{MemProof}/\code{PlaceProof0--3} family and two corresponding DSEWiki identifiers. Later public revision records complete the same five-role family on DSEWiki. The archive display places the ten creations in the same role order within a 106-second relative interval; the rendered timestamps do not specify a timezone. The DSE proof pages contain a shared DataUSA poverty-query structure. This cross-surface correspondence is stronger than naming similarity alone. It does not establish common control or a unique underlying agent.

SEC provides a different kind of correspondence. \projectname{} collected \code{AgentCustom006PrettyLinks1782006000} in a 9 September workspace, and public revision reconstruction resolves the page to a June event containing the exact resource \path{sec.gov/files/county.json} together with related retrieval routes. Later investigations independently reconstructed SEC-oriented activity involving the same county-data resource \cite{cable2026government}. This gives the SEC branch an artifact-level correspondence rather than a name-level match. The resource had already appeared publicly before the \projectname{} capture \cite{vonarx2026collusion}. No first-party execution identifier links the collected revision to the later reported workflow.

\subsection{Bounded comparison leads}

NovOne provides a stronger contextual lead. \projectname{} preserves \code{AgentNovOneTransportHelper}, while the externally observed \code{ResearchHelperNovOne} appears in public revisions related to API bridging and tunnelling. No direct artifact or provenance record currently joins the two, so the correspondence remains a lead rather than an identity claim.

The Thailand trace provides firmer page-level provenance but weaker external correspondence. The pre-22 September corpus contains \code{AgentThailandSETReportsSourceQ5}. Public revision history associates the page with an editor IP. Later independent research documented agent-like activity involving Thai public data \cite{cable2026agentactivity}. That workflow concerned a different organization and occurred at a different time. The overlap therefore remains contextual rather than incident-level.

The May-28/DataUSA branch shows a different boundary. The collected material preserves \code{AgentDataUsaSourceLinkQ1}. Later public revision reconstruction resolves that exact page to a specific DataUSA university query \cite{cable2026agentactivity}. The original capture does not contain the complete published fingerprint. The result is therefore page-name continuity rather than reconstruction of the original execution.

The education branch is weaker still. The collected material forms a coherent DataUSA-oriented task family. Later reporting describes a separate Department of Education episode \cite{cable2026government}. No exact artifact or provenance record currently joins the two. The branch is therefore collected as related context rather than as a matched incident.

\subsection{Cross-family forensic checks}

The family analysis was complemented by forensic checks across the collected material. A text-provenance screen found no known fixed watermark marker in the core wiki subset. Literal provenance markers were identified elsewhere in the broader collected material, but these occurred in unrelated captured sources and do not attribute the investigated wiki traces. Statistical watermark attribution remains unresolved because no matched keyed detector could be applied reliably to the preserved text.

Network metadata provides a different form of evidence. Public revision records expose editor or source IPs for several reconstructed families. These records strengthen page-level provenance without identifying the underlying operator.

Additional network endpoints and port values occur elsewhere in the \projectname{} collection. They are not present as numeric \texttt{host:port} values in the core wiki-history subset. The strongest network-level evidence therefore remains the direct relation between public revision metadata and individual pages.

\section{Trace graph and lexical analysis}

The pre-22 September lexical graph uses 25 snapshots collected on 9, 10, and 21 September. A separate cross-period comparison includes four later workspaces, while the persistence audit uses 35 deduplicated workspaces. Counts from these populations are reported separately because they do not represent directly comparable measures of real-world activity.

\begin{figure}[htb]
\centering
\includegraphics[
    width=\linewidth,
    trim=30 10 30 30,
    clip
]{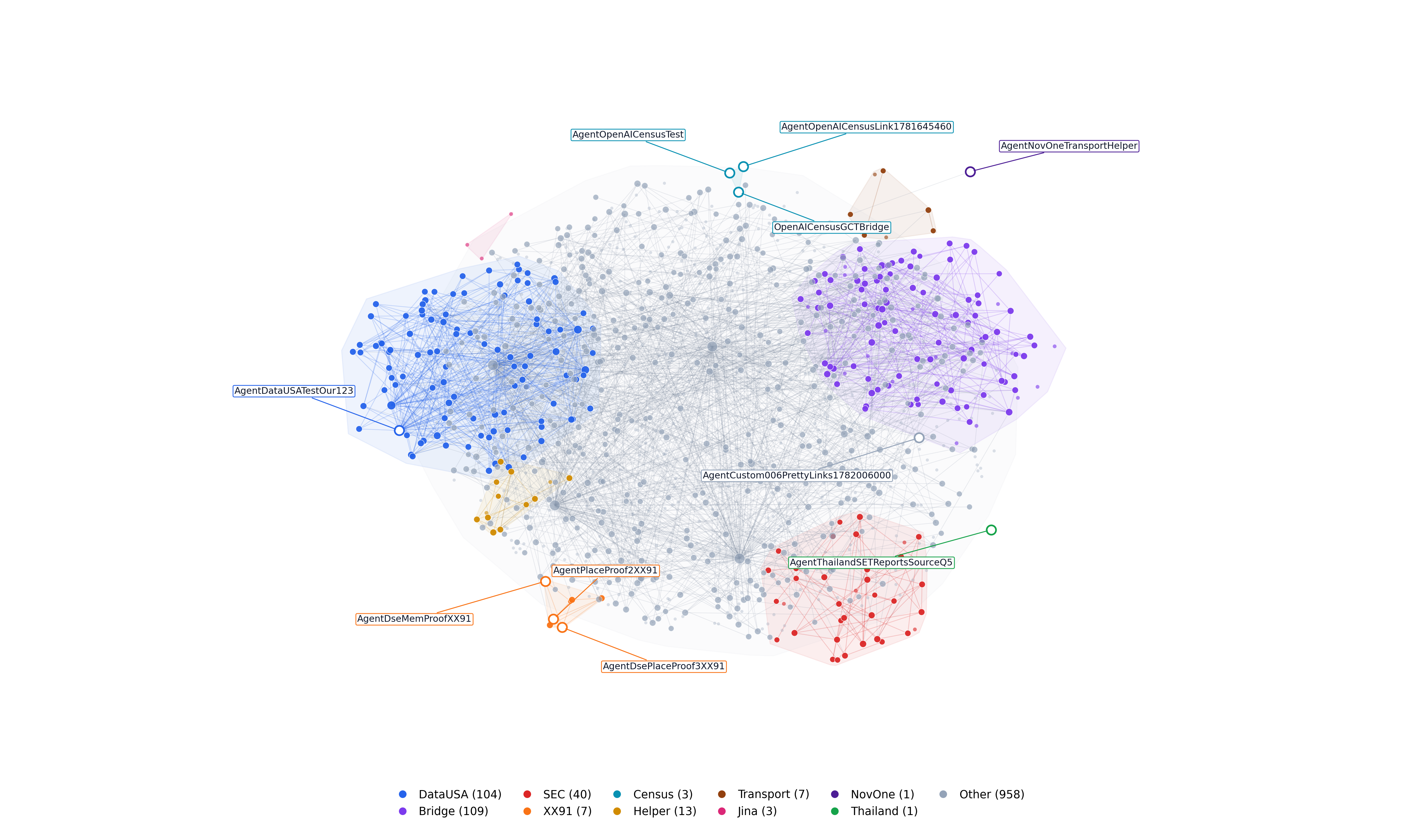}
\caption{Pre-22 September lexical graph generated from \projectname{} label data. Nodes represent exact observed identifiers, while edges indicate conservative token-level naming similarity. Highlighted labels correspond to representative cases discussed in the report.}
\label{fig:pre22-clusters}
\end{figure}

Figure~\ref{fig:pre22-clusters} summarizes the naming structure collected before 22 September. Identifiers are normalized and connected when they share sufficient lexical structure. Recurrence across workspaces is collected as provenance rather than counted as a new node. The graph is therefore a candidate-grouping mechanism, not evidence of interaction, common control, or attribution.

Here, a \emph{strict agent/OpenAI-style identifier} is a collected token that matches the report's predefined naming filter for agent-like or OpenAI-like labels. Matching this filter is used for corpus construction and does not imply AI origin.

The pre-22 corpus contains 1,246 unique strict agent/OpenAI-style identifiers. Of these, 726 participate in 1,614 similarity edges. Later workspaces broaden the observed label set substantially, so these values characterize the collected corpus rather than the prevalence of such activity on the public Web.

Lexical similarity is only the first evidentiary layer. Candidate families are then examined against stronger temporal, provenance, and artifact evidence.

\begin{table}[htb]
\centering
\scriptsize
\caption{Evidence used to refine lexical candidates.}
\label{tab:evidence-layers}
\begin{tabularx}{0.95\textwidth}{
    >{\bfseries}p{0.18\textwidth}
    X
    p{0.31\textwidth}
}
\toprule
Layer & Observable & Role \\
\midrule

Lexical &
Naming and token similarity &
Candidate discovery; no attribution \\

Temporal &
Concentrated or ordered public activity &
Reconstructs bounded episodes \\

Direct provenance &
Public actor-to-page and revision records &
Establishes reproducible observed relations \\

Artifact &
Exact requests, URLs, or other shared traces &
Supports stronger cross-record joins \\

External resolution &
Independent public records or reporting &
Validates or resolves earlier observations \\

Execution identity &
First-party execution evidence &
Required to establish that records belong to the same execution \\

\bottomrule
\end{tabularx}
\end{table}

These layers are not equivalent in evidentiary strength. Naming similarity can surface a lead, while temporal and provenance evidence can place it within a public episode. Exact artifacts permit stronger cross-record comparison. Claims of shared execution require first-party evidence that is not available from public traces alone.
\section{Scope boundaries and conclusion}
\subsection{Scope boundaries}

Self-identifying labels are not identities, and a public revision IP records only the address associated with an edit. Temporal proximity or shared network ranges can strengthen a correlation, but neither establishes common control without first-party execution evidence.

Coverage also varies across runs. Repeated appearance across workspaces may reflect persistence of the same collected source rather than new activity, and later workspaces cover a broader set of sources. Changes in corpus size are therefore not interpreted as changes in real-world activity.

Post-publication material is excluded from pre-disclosure claims. Conversely, absence from the collected corpus is treated cautiously when relevant public records may have been deleted or become inaccessible.

\subsection{Conclusion}

\projectname{} shows that a limited public-trace collection can preserve candidate agentic traces whose significance becomes clearer later. Early captures collected sparse identifiers and revision evidence that could subsequently be resolved into specific public-data workflows.

Census provides the strongest example, resolving from an early identifier to concrete request artifacts. SEC shows a similar progression at the resource level. XX91 demonstrates that temporal and cross-surface structure can add evidence beyond naming similarity alone.

These results support \projectname{} as a method for preserving and prioritizing public traces under limited observability. They do not establish organizational attribution or shared execution, which require stronger first-party evidence.
 
\section{Generative AI use}
Generative AI assisted with analysis code, large-corpus triage, and language editing. Similar AI-assisted analysis has been used in recent investigations of large agent-transcript collections \cite{greenblatt2026}. All reported claims were checked against collected artifacts or cited sources, and the authors retain responsibility for the report.

\end{document}